\documentclass[aps,prb,groupedaddress,showpacs]{revtex4}

\begin{document}


\title{Coherence properties of bulk matter}


\author{Marco Frasca}
\email[]{marcofrasca@mclink.it}
\affiliation{Via Erasmo Gattamelata, 3 \\ 00176 Roma (Italy)}


\date{\today}

\begin{abstract}
We prove a theorem, using the density functional approach and relying on a classical
result by Lieb and Simon on Thomas-Fermi model, showing that in the thermodynamic
limit bulk matter is at most semiclassical and coherence preserving. The connection
between quantum fluid dynamics and density functional theory in the formulation due
to Kohn and Sham plays a significant role leading to a Vlasov-Poisson system of equations
for the Wigner function. Coherence stability is achieved by noting that small oscillations
in bulk matter are damped by Landau damping. In some conditions the 
initial Wigner function could generate an opposite effect and
coherence stability can be lost involving higher order quantum effects
for a macroscopic body. 
\end{abstract}


\maketitle

\section{Introduction}

The proof of the stability of the matter, initially due to Dyson and Lennard \cite{dl1,dl2}
with a mathematical tour de force, and largely improved by an elegant argument due
to Lieb and Thirring \cite{lt}, is surely one of the most important results about the
existence of bulk matter. The situation is far less clear about the coherence properties
of matter, the aim of this paper.

The problem of addressing a many-body system from a quantum mechanical standpoint is
rather demanding and difficult to treat unless the number of components is really small.
A smart approach to this problem arose initially with the pioneering work by Thomas and
Fermi \cite{th,fe} where was shown how density is all one needs to address the physics of
large atoms, then theorems by Kohn and Hohenberg \cite{kh} and Kohn and Sham \cite{ks} 
made this approach mature to analyze the behavior of large systems relying just on 
the computation of the density. Striking agreement between experimental and
computed data is so achieved by this so called density functional theory (DFT).

One of the main drawbacks of this approach, notwithstanding the great successes, should
be identified with the impossibility to analyze the coherence properties of bulk matter.
A question like ``is bulk matter a classical or a quantum object?'' 
or ``does bulk matter preserve coherence?'' cannot be addressed
just considering density.

Indeed, these questions can be addressed inside the same theory by a set of results
recently obtained \cite{mk}. A Wigner function can be associated to a density functional and the
coherence properties of bulk matter can be analyzed in the thermodynamic limit.

A relevant result about a many-body quantum theory was obtained by Lieb and Simon\cite{ls1,ls2}
giving to Thomas-Fermi theory a role of limit theory for the quantum counterpart when the
number of particles is taken to go to infinity, the limit that is of interest for our aims.
Thomas-Fermi theory also played a relevant role in the argument of Lieb and Thirring on 
stability of matter. Then, this theory should be considered a substantial aspect of
many-body physics.

Being Thomas-Fermi theory a semiclassical theory, indeed is the leading order of a 
Wigner-Kirkwood expansion of the Green function of the quantum system\cite{nt}, 
we can conclude that bulk
matter, in the limit of a large number of particles, has classical properties that
can be preserved during time evolution. This conclusion about stability is obtained by
analyzing the time evolution of the Wigner function obtained by time dependent DFT plus
the Lieb and Simon theorem. Coherence stability means that small perturbations on the
system are damped. This is a property of the Vlasov-Poisson equations that drive the
system in this case \cite{nt}. On the other hand, if matter is initially properly prepared,
coherence stability can be lost and small perturbations can be amplified. Then,
classicality, a common property, can be lost.

The paper is so structured. In sec.\ref{sec2} we introduce the model and, by using the semiclassical Wigner-Kirkwood 
expansion and the Lieb and Simon theorem, we show that the limit of the quantum theory for the number of particles
going to infinity and $\hbar\rightarrow 0$ coincide with the Thomas-Fermi model. 
In order to study coherence properties of bulk matter we need a time-dependent model. 
This is obtained in sec.\ref{sec3} by the Runge and Gross theorem giving the time-dependent density functional 
theory that translate into a hydrodynamic model for the Thomas-Fermi model. 
Having a hydrodynamic model, a kinetic equation for the Wigner function is obtained in 
sec.\ref{sec4} and, by a recently proved theorem, this is equivalent to the classical limit 
plus a quantum correction giving the full set of equations known as Vlasov-Poisson equations. 
This set can produce for small perturbations a damping granting coherence stability. 
Exotic initial Wigner functions can generate instabilities. In sec.\ref{sec5} the conclusions are given.

\section{\label{sec2} Thomas-Fermi Theory as Semiclassical Limit of a Quantum Theory}


All the low energy phenomenology is fully described by the Hamiltonian \cite{lau}
\begin{equation}
    H = -\sum_{j=1}^{N_e}\frac{\hbar^2}{2m}\nabla^2_j
	    -\sum_{\alpha=1}^{N_i}\frac{\hbar^2}{2M_\alpha}\nabla^2_\alpha
		-\sum_{j=1}^{N_e}\sum_{\alpha=1}^{N_i}\frac{Z_\alpha e^2}{|{\bf x}_j-{\bf R}_\alpha|}
		+\sum_{j<k}\frac{e^2}{|{\bf x}_j-{\bf x}_k|}
		+\sum_{\alpha<\beta}\frac{Z_\alpha Z_\beta e^2}{|{\bf R}_\alpha-{\bf x}_\beta|}
\end{equation}
that is able to describe all the properties of matter as far as we know. We have
put $N_e$ the number of electrons, $N_i$ the number of positive ions, $e$ the electron charge,
$Z_\alpha$ the number of positive charges for each ion.

Introducing the density and neglecting for our aims the repulsion between positive ions
assumed to be pointlike and classical, 
the quantum system we are going to consider is rather well-known,
being the starting point of a lot of low energy condensed matter analysis \cite{par},
having the following energy functional:
\begin{equation}
\label{eq:H}
    {\cal E} = T - Ze^2\int\frac{\rho(\bf x')}{|\bf x'|}d^3x'+
        \frac{e^2}{2}\int\int\frac{\rho(\bf x')\rho(\bf x'')}{|\bf x' - \bf x''|}d^3x'd^3x''
\end{equation}
being $T$ the kinetic energy
given by
\begin{equation}
    T = \int d^3x_1d^2x_2d^3x_3\ldots d^3x_N\Psi^*({\bf x_1,\bf x_2,\ldots,\bf x_N})
	\left(-\frac{1}{2m}\sum_{i=1}^N\Delta_{2i}\right)\Psi(\bf x_1,\bf x_2,\ldots,\bf x_N)
\end{equation}
being $m$ the electron mass,
$\rho(\bf x)$ the electronic density computed by the Slater
determinant $\Psi$, 
if the Hartree-Fock approximation is invoked,
as $\int d^2x_2d^3x_3\ldots d^3x_N\Psi^*({\bf x,\bf x_2,\ldots,\bf x_N})
\Psi(\bf x,\bf x_2,\ldots,\bf x_N)$,
$Z$ is the number of positive charges and $e$ the electron charge. We assume neutrality, that is,
the number of electrons, $N=\int\rho({\bf x})d^3x$, 
is the same as the number of positive charges in the system.
One could transform the functional ${\cal E}$ into a density functional ${\cal E}[\rho(\bf x)]$,
bypassing self-consistent Hartree-Fock equations
for the Hartree-Fock approximation,
if it would be possible to obtain the kinetic energy as a functional of $\rho(\bf x)$.

We would like to accomplish such a task in the classical limit $\hbar\rightarrow 0$. We can 
reach our aim by referring to standard results in nuclear physics \cite{nt} and Bose 
condensation \cite{str} as
also given in \cite{par}: {\sl The density matrix can be developed into a $\hbar\rightarrow 0$
series, derived from the Wigner-Kirkwood series for the Green function 
for a single particle moving in the Hartree-Fock potential $V$\cite{sch}}
\begin{equation}
  G({\bf x},{\bf x'};t)=G_{TF}({\bf x},{\bf x'};t)
\left[1+\frac{\hbar^2}{12m}\left(\frac{t^2}{\hbar^2}\Delta V 
- i\frac{t^3}{\hbar^3}|\nabla V|^2\right)+\ldots\right],
\end{equation}
{\sl whose leading order is just the Thomas-Fermi approximation }
\begin{equation}
G_{TF}({\bf x},{\bf x'};t)=\left(\frac{m}{2\pi i\hbar t}\right)^{\frac{3}{2}}
\exp\left[\frac{im({\bf x}-{\bf x'})^2}{2\hbar t}
-\frac{it}{\hbar}V\left(\frac{{\bf x}+{\bf x'}}{2}\right)\right]
\end{equation}
{\sl giving the density of kinetic energy as}
\begin{equation}
    \tau_{TF} = \frac{3}{10}\frac{\hbar^2}{m}(3\pi^2)^{\frac{2}{3}}\rho({\bf x})^{\frac{5}{3}}.
\end{equation}
So, we can conclude that the Thomas-Fermi approximation can represent classical objects being the
leading order of a semiclassical approximation.

For our aims this is not enough as we are interested in the limit of a large number of particles 
in the energy functional we started with. Indeed, there is another theorem due to Lieb and 
Simon\cite{ls1,ls2} that gives an answer to this question: 
{\sl The limit of number of particles $N$ going to infinity for energy functional (\ref{eq:H}) 
is the Thomas-Fermi model}. This means that
in this limit we recover again the leading order of a semiclassical approximation. So,
classical objects can be obtained when the number of particles in a system increases without bound
because {\sl the limit of a number of particles going to infinity coincides with the semiclassical
limit $\hbar\rightarrow 0$}. This is the main result of this section.

One can ask at this point if this property is stable, that is, if classicality is indeed
preserved during time evolution or if some instabilities may set in moving the system
toward some generic quantum state. In the following we will answer to this question
relying on the results of this section but to do this we need to study the time evolution
of the quantum system. We develop this analysis in the next section.

\section{\label{sec3} DFT and Quantum Fluid Dynamics}

In the preceding section we have seen that the energy functional can be turned into a density functional. 
Indeed, Thomas-Fermi theory turns out to be the simplest of density functional theories and the 
relevance of this approach relies on a couple of theorems proved by Kohn and Hohenberg \cite{kh}. 
On the basis of these results the density $\rho(\bf x)$ contains all the
informations about the ground state and there is no need to have the full wave function to
get the properties of the system (first theorem). Once one has a density functional, a variational principle
gives the equations for the computation of the density (second theorem). This is generally done numerically by solving
a set of self-consistent single particle equations as devised by Kohn and Sham \cite{ks}.
Hohenberg-Kohn theorems have been extended to the time-dependent case by Runge and Gross \cite{rg} but now the 
variational principle changes into a stationary point principle for
the functional
\begin{equation}
   {\cal A}(t)=\int_{t_0}^{t_1}\langle\Psi(t)|i\hbar\partial/\partial t - H(t)|\Psi(t)\rangle.
\end{equation}

All these results apply straightforwardly to the simplest case of the Thomas-Fermi approximation 
producing a hydrodynamic model \cite{drs} that can be cast into the form
\begin{eqnarray}
    \frac{\partial\rho}{\partial t}&+&\nabla\cdot(\rho {\bf u})=0 \\
    m\frac{\partial\chi}{\partial t}+\frac{m}{2}(\nabla\chi)^2&+&
    \frac{\delta U}{\delta\rho} + \frac{\delta T}{\delta\rho} = 0
\end{eqnarray}
being ${\bf u}=\nabla\chi$,
\begin{equation}
    U[\rho({\bf x},t)] = - Ze^2\int\frac{\rho({\bf x'},t)}{|\bf x'|}d^3x'+
        \frac{e^2}{2}\int\int\frac{\rho({\bf x'},t)\rho({\bf x''},t)}{|\bf x' - \bf x''|}d^3x'd^3x''
\end{equation}
and
\begin{equation}
    T[\rho({\bf x},t)] = \int d^3x\frac{3}{10}\frac{\hbar^2}{m}(3\pi^2)^{\frac{2}{3}}
    \rho({\bf x},t)^{\frac{5}{3}}.
\end{equation}
To get this equations we have used the ansatz for the signle particle Kohn-Sham orbital
$\phi_i({\bf x},t)=\phi_i^0({\bf x},t)\exp\left[i\frac{\hbar}{m}\chi({\bf x},t)\right]$
where we have assumed a single phase for each orbital. As we will see this approximation
simplifies the computations but does not invalidate our conclusions.

We can now give explicitly the above set of equations through the density and the velocity
field. One has
\begin{equation}
    m\frac{\partial {\bf u}}{\partial t}+ m ({\bf u}\cdot\nabla){\bf u}+
    \nabla V + \frac{1}{\rho}\nabla P_F = 0
\end{equation}
being $V$ the potential solution of the Poisson equation
\begin{equation}
    \Delta_2 V({\bf x},t) = 4\pi e^2 [\rho({\bf x},t)-Z\delta({\bf x})]
\end{equation}
and $P_F$ the Fermi pressure given by
\begin{equation}
    P_F({\bf x},t) = \frac{\hbar^2}{5m}(3\pi^2)^{\frac{2}{3}}\rho({\bf x},t)^{\frac{5}{3}}.
\end{equation}
that contains all the quantum information. We have obtained a time-dependent Thomas-Fermi model expressed through
Euler hydrodynamic equations. This set of equations can be transformed into a single Schr\"odinger equation\cite{drs}
but we do not pursue this matter here.

It is important to notice that the simple knowledge of the density it is not enough to get informations on
the coherence properties of a many-body quantum system. But, having introduced hydrodynamic equations by the
density, we can associate to it a Wigner function $W({\bf x},{\bf p},t)$ in the phase space containing all the
informations about the evolution of the system. It is important to notice at this point that, because the Thomas-Fermi
model corresponds to Euler equations, no collisional term does enter into the kinetic equation for the Wigner
function. This opens the way to the main result of this paper, that is, classicality, when emerges, is a stable
property of a many-body quantum system being the system, in the limit of a large number of particles, driven by
a classical set of kinetic equations, known in the current literature as the Vlasov-Poisson equations.

\section{\label{sec4} Quantum Fluid Dynamics and Wigner Function}
For our aim, the set of hydrodynamic equations can be cast in the form
\begin{eqnarray}
    \frac{\partial\rho}{\partial t}&+&\nabla\cdot(\rho {\bf u})=0 \\
	\frac{\partial {\bf u}}{\partial t}&+& ({\bf u}\cdot\nabla){\bf u}+
    \frac{1}{m}\nabla (V + {\cal E}_F) = 0 \\
	\Delta_2 V({\bf x},t) &=& 4\pi e^2 [\rho({\bf x},t)-Z\delta({\bf x})]
\end{eqnarray}
being ${\cal E}_F=\frac{\hbar^2}{2m}(3\pi^2)^{\frac{2}{3}}\rho({\bf x},t)^{\frac{2}{3}}$ the
Fermi energy. By a recently proved theorem due to J\"ungel and Wang\cite{jw} this set of equations derives from the 
self-consistent set of Wigner-Poisson equations
\begin{eqnarray}
	\frac{\partial W}{\partial t}&+&  {\bf v}\cdot\nabla_x W-
    \frac{1}{m}\nabla_x (V + {\cal E}_F)\cdot\nabla_v W= 0 \\
	\Delta_2 V({\bf x},t) &=& 4\pi e^2 [\rho({\bf x},t)-Z\delta({\bf x})] \\
	\rho({\bf x},t) &=& \int W({\bf x},{\bf v},t)d^3v
\end{eqnarray}
being $W({\bf x},{\bf v},t)$ the Wigner function.
This last set of equations is all one needs to describe low energy phenomenology
of bulk matter in the limit of number of particles going to infinity.
The set we have obtained is the Vlasov-Poisson set representing the leading order of a 
semiclassical approximation to the full Wigner equation\cite{jw}. We note that we have to keep a quantum 
term due to the Pauli principle, that is, the fermionic nature of the matter. 
This shows one of the main results of the paper, that is, when the number of particles goes 
to infinity the coherence is ruled by a Vlasov-Poisson set of equations representing the classical limit 
of the Wigner equation but with a quantum correcting term. 
The correcting term is just the Fermi energy that for a macroscopic body is negligible with respect 
to the self-consistent potential. A similar approximation is generally assumed in Bose condensates for the 
Thomas-Fermi approximation to the Gross-Pitaeskvii equation\cite{str}.

This permits us to get finally the proof of our theorem: {\sl classicality as an emerging property of a 
macroscopic body is a stable property}. This can be seen easily assuming an initial Wigner function, 
as the one derived from the Fermi-Dirac distribution at zero temperature, that warrants Landau damping 
of small perturbations \cite{pla} and taking into account neutrality of the matter. It is interesting to 
note that some possible initial conditions may generate unstable macroscopic bodies having quantum properties 
as damping can change into an exponential increase of the perturbation and stability could just be reached 
by properly including higher order quantum effects.

In order to see the above conclusions at work, we analyze the case of an isotropic superconductor.
This implies a constant wavefunction, when no current flows, that in turn means that the
initial Wigner function will be proportional to $\delta(v)$, being $v$ the modulus of the velocity. 
One can prove from the theory of Landau damping\cite{pla} that the condition for a small
initial oscillation to die out is
\begin{equation}
    \left(\frac{\partial\tilde W_0(k,v)}{\partial v}\right)_{v=\frac{\omega}{k}}<0
\end{equation}
being $\tilde W_0(k,v)$ the initial Wigner function after a Fourier transform in space.
So in this case the Landau damping factor is null. Indeed,
this initial Wigner function is well-known in literature for the Vlasov-Poisson equations giving
rise to instability \cite{man} and so, in this, higher order quantum corrections are needed to make
the system stable. On the other side, if we take a Fermi distribution function in the limit of
high temperature, this reduces to a Boltzmann distribution
giving rise to Landau damping proper to a stable classical system.

\section{\label{sec5} Conclusions}

By a sequence of known results we have shown how coherence can be considered for a many-body quantum system. 
Classicality proves to be an emerging property in the limit of a large number of particles as this limit is 
given by the Thomas-Fermi model as also happens to the semiclassical limit $\hbar\rightarrow 0$ at the leading order. 
Evolution in time of the density functional, provided by Runge and Gross theorem translates into a set of 
hydrodynamic equations for the Thomas-Fermi model. These equations are equivalent to the
set of Vlasov-Poisson kinetic equations for the Wigner function, corresponding to the classical 
limit plus a quantum correction given by the Fermi energy, that gives rise to Landau damping of small 
perturbations assuring stability of classicality in the considered limits.

\end{document}